\documentstyle[prd,aps]{revtex}
\begin{document}
\draft
%%%%%%%%%%%%%%%%%%%%%%%%%%%%%%%%%%%%%%%%%%%%%%%%%%%%%%%%%%%%%%%%%%
%
%  Uncomment following two lines and one below for 2 column format.
%
\twocolumn[\hsize\textwidth\columnwidth\hsize\csname
@twocolumnfalse\endcsname
%%%%%%%%%%%%%%%%%%%%%%%%%%%%%%%%%%%%%%%%%%%%%%%%%%%%%%%%%%%%%%%%%%
\preprint{SUSSEX-AST 96/7-2, astro-ph/9607038}
\title{Four-year COBE normalization of inflationary cosmologies}
\author{Emory F. Bunn}
\address{Department of Astronomy and Astrophysics, University of
California, Berkeley, California 94720}
\author{Andrew R. Liddle}
\address{Astronomy Centre, University of Sussex, Falmer, 
Brighton BN1 9QH, United Kingdom} 
\author{Martin White}
\address{Enrico Fermi Institute, University of Chicago, 5640 S.~Ellis
Ave, Chicago, Illinois 60637}
\date{\today}
\maketitle
\begin{abstract}
We supply fitting formulae enabling the normalization of slow-roll
inflation models to the four-year COBE data. We fully include the
effect of the gravitational wave modes, including the predicted relation of
the amplitude of these modes to that of the density perturbations. We 
provide the normalization of the matter power spectrum, which can be 
directly used for large-scale structure studies. The normalization for 
tilted spectra is a special case. We also provide fitting functions for the 
inflationary energy scale of COBE-normalized models and discuss the validity 
of approximating the spectra by power-laws. In an Appendix, we extend our 
analysis to include models with a cosmological constant, both with and 
without gravitational waves.
\end{abstract}
\pacs{PACS numbers: 98.70.Vc, 98.80.Cq \hspace*{0.7cm} Sussex 
preprint SUSSEX-AST 96/7-2,  astro-ph/9607038}

%  This is the other line to be uncommented for 2 column format
\vskip2pc]
%%%%%%%%%%%%%%%%%%%%%%%%%%%%%%%%%%%%%%%%%%%%%%%%%%%%%%%%%%%%%%%%%%

\section{Introduction}

The four-year data set from the Cosmic Background Explorer (COBE)
satellite \cite{COBE4} is the last word we shall hear for some time
concerning irregularities on the largest observable scales. One of the
most important uses of the COBE data is in providing an accurate
normalization of the power spectrum of density perturbations for a
given set of theoretical assumptions; once normalized, one can then
compare the theory to a wide range of observations of large-scale
structure in the Universe.

The aim of this short paper is to apply the techniques of Bunn and White
\cite{BW} to normalize slow-roll inflation models. As well as a spectrum of 
density perturbations, inflationary models predict a spectrum of 
gravitational waves, which can influence the large-angle microwave 
anisotropies seen by COBE. Further, in inflationary models one predicts a 
generic link between the density perturbations and gravitational waves, 
which must be taken into account in obtaining an accurate normalization. In 
this paper, we provide a self-contained account of how to predict
these spectra from an inflationary model and quote fits to COBE for the
normalization of the matter power spectrum, which can be directly used
for large-scale structure studies. We also give fitting functions for the 
inflationary energy scale.

\section{Spectra from Inflation}

We are of course unable to provide information for an arbitrary
inflationary model. We shall not consider models with more than one
dynamical field, for which the calculations are extremely involved 
\cite{2field} and for which initial conditions may be important, and we 
are also unable to consider open inflationary models \cite{open} because in 
that case no-one has yet managed to compute the gravitational wave spectrum. 
We therefore restrict ourselves to the usual single-field inflation models, 
driven by a rolling scalar field $\phi$, known generically as chaotic 
inflation \cite{chaotic}. This situation is a very general one, because most 
two-field models, for example hybrid inflation \cite{hybrid}, feature only a 
single dynamical scalar field, and models with extended gravity sectors can 
usually be brought into the Einstein form via conformal transformation 
\cite{conformal}. 

Exact results for the spectra are not known for arbitrary potentials
$V(\phi)$. They can be calculated analytically via the slow-roll
approximation. The accuracy required depends on the way in which the
results shall be used, and for normalizing to COBE it is valid to use
the well known lowest-order results (see, e.g., Ref.~\cite{LLrep}), which
give the density perturbation (i.e.~scalar) spectrum $A_{{\rm S}}(k)$
and gravitational wave (i.e.~tensor) spectrum $A_{{\rm T}}(k)$ as
\begin{eqnarray}
\label{dp}
A_{{\rm S}}^2(k) & = & \frac{512\pi}{75} \, \left. 
	\frac{V^3}{m_{{\rm Pl}}^6 V'^2} \right|_{k=aH} \,,\\
A_{{\rm T}}^2(k) & = & \frac{32}{75} \, \left. \frac{V}{m_{{\rm Pl}}^4} 
	\right|_{k = aH} \,,
\end{eqnarray}
where prime indicates derivative with respect to $\phi$,
and the right-hand side is to be evaluated at the $\phi$ value when
the scale $k$ equals the Hubble scale during inflation.  The precise
definition of the spectra will be clarified later.

More accurate expressions than these do exist in the literature \cite{SL}.
However, these are {\em not} necessary for discussion of the COBE 
normalization, though they will be needed for discussion of the inflationary 
energy scale, as discussed later.

The slow-roll approximation is characterized by the smallness
(relative to unity) of two parameters\footnote{Beware that two
slightly different versions of these exist in the literature,
depending on whether the fundamental quantity is taken to be the
potential or the Hubble parameter \cite{LPB}. We use the former.} 
\cite{LL,LLrep}
\begin{equation}
\epsilon \equiv \frac{m_{{\rm Pl}}^2}{16\pi} \left( \frac{V'}{V}
	\right)^2 \quad , \quad \eta \equiv 
	\frac{m_{{\rm Pl}}^2}{8\pi} \, \frac{V''}{V} \,.
\end{equation}
Using Eq.~(\ref{efold}) below, the spectral indices of the two spectra
can be written in terms of these parameters
\begin{eqnarray}
n - 1 & \equiv & \frac{d \ln A_{\rm S}^2(k)}{d \ln k} = 
	- 6 \epsilon + 2 \eta \,, \\
n_{{\rm T}} & \equiv & \frac{d \ln A_{{\rm T}}^2(k)}{d \ln k} = - 2 
	\epsilon \,,
\end{eqnarray}
as can the ratio of the two spectra
\begin{equation}
\label{relamp}
\frac{A_{{\rm T}}^2}{A_{\rm S}^2} = \epsilon = - \frac{n_{{\rm T}}}{2} \,.
\end{equation}
Note this ratio is not independent of the tensor spectral index; the 
gravitational wave and density perturbation spectra are related due to their 
common origin in a single potential $V(\phi)$ \cite{LL}.\footnote{Only the 
spectra are related. The phases within a given realization are 
uncorrelated.}

All the above expressions apply at any scale $k$, with the
spectral indices able to vary with scale. Since COBE covers a
fairly restricted range of scales, the spectra produced can be
approximated by power-laws. Then we need to specify the amplitudes and
spectral indices only at a single scale. It is best to choose this
scale near the center of the COBE data, so we evaluate them at the scale
$k_* = 7 a_0 H_0$. Since, loosely speaking, the $\ell$-th microwave
multipole samples scales around $k = \ell a_0 H_0/2$, this corresponds 
to the fourteenth multipole. Throughout, subscript `*' will indicate 
evaluation at this scale, and subscript `0' indicates present value, 
here of the scale factor $a$ and Hubble parameter $H$.

For a specification of the COBE normalization to have a precise meaning, we 
need to take care in relating scales during inflation to present scales. The 
number of $e$-foldings $N$ before the end of inflation at which $k=aH$ 
is given by (see, e.g., Ref.~\cite{LLrep})
\begin{eqnarray}
\label{scales}
N(k) & = & 62 - \ln \frac{k}{a_0 H_0} - 
	\ln \frac{10^{16} {\rm GeV}}{V_k^{1/4}} \nonumber \\
 & & \quad \quad \quad \quad + \ln \frac{V_k^{1/4}}{V_{{\rm end}}^{1/4}} 
	- \frac{1}{3} \ln \frac{V_{{\rm end}}^{1/4}}{\rho_{{\rm reh}}}
	\,.
\end{eqnarray}
Here $V_k$ is the potential when $k=aH$, $V_{{\rm end}}$ is the potential
at the end of inflation and $\rho_{{\rm reh}}$ is the energy density
immediately after reheating has completed, resuming standard big bang
evolution.

The appropriate point on the inflationary potential to evaluate the 
spectra is given by $N_*$. It depends on the energy scale of inflation, 
which itself depends on the normalization, fortunately only weakly. Once 
the COBE normalization is found, $V_*$ and $V_{{\rm end}}$ are 
determined in the context of a specific model, and the normalization can be 
iteratively improved if desired to take these values into account. 
However, the reheat energy is much more uncertain, and so consequently $N_*$ 
is not normally specified very accurately. Often, $N_*$ is taken to be 
60 or 50.

To locate the $\phi$-value when $k_* = aH$ during inflation, one simply 
carries out the integral
\begin{equation}
\label{efold}
N(\phi) \simeq \frac{8 \pi}{m_{{\rm Pl}}^2} \int_{\phi_{{\rm 
	end}}}^{\phi} \frac{V}{V'} \, d\phi \,,
\end{equation}
where $\phi_{{\rm end}}$ could be calculated numerically but is
normally given to adequate accuracy by the breakdown of the slow-roll
conditions, taken as $\epsilon_{{\rm end}} = 1$.\footnote{In models where
inflation doesn't end by steepening of the potential, such as hybrid
inflation \cite{hybrid}, $\phi_{{\rm end}}$ is given by a different
condition such as an instability condition.}
Having located $\phi_*$, one immediately gets the slow-roll parameters
and hence the spectral indices at that scale. The tensor spectral
index $n_{{\rm T}}$, which is the hardest thing to directly observe,
can be eliminated through its relation to the ratio
$A_{{\rm T}}^2/A_{{\rm S}}^2$. To indicate the amount of tensors, we
define a quantity $r$ by \cite{LL,LLrep}
\begin{equation}
\label{tensca}
r = 12.4 \, \frac{A_{{\rm T}}^2 (k_*)}{A_{{\rm S}}^2(k_*)} \,,
\end{equation}
which measures, in the matter-dominated and Sachs-Wolfe approximations, the 
relative importance of gravitational waves and density perturbations in 
contributing to the relevant microwave multipole, in this case the 
fourteenth. Henceforth, $n$ and $n_{{\rm T}}$ will also be assumed to be 
evaluated at $k_*$.

\section{Fitting to the four-year COBE data}

\subsection{Normalization of the power spectrum}

Large-scale structure studies require the normalization of the present-day 
power spectrum. We precisely define our notation for the initial spectrum 
$A_{{\rm S}}(k)$ here. In a critical-density universe, it is related to the 
rms fluctuation per logarithmic $k$-interval $\Delta^2(k)$, and to the
usual power spectrum $P(k)$, both at the present epoch, by
\begin{equation}
\label{powerspec}
\Delta^2(k) \equiv \frac{k^3 P(k)}{2\pi^2} \equiv 
	\left(\frac{k}{a_0 H_0}\right)^4 \,
	A_{{\rm S}}^2(k) \, T^2(k) \,,
\end{equation}
where $T(k)$ is the usual transfer function, normalized to unity on
large scales. The variance of the density field, smoothed on scale
$R$, is given by
\begin{equation}
\sigma^2(R) = \int \frac{dk}{k} \, \Delta^2(k) \, W^2(kR) \,,
\end{equation}
where the smoothing function $W(kR)$ tends to unity at small $k$.
The observables related to $A_{{\rm T}}$ are discussed in 
Refs.~\cite{TurWhi}.

The fitting to COBE is described in Ref.~\cite{BW}. Rather than use 
Eq.~(\ref{relamp}) to set the relative normalization of scalars and tensors, 
we use the more accurate expression
\cite{CKLL2,TurWhi}
\begin{equation}
\frac{A_{{\rm T}}^2(k_*)}{A_{{\rm S}}^2(k_*)} = - \frac{n_{{\rm T}}}{2}
	\left( 1 - \frac{n_{{\rm T}}}{2} + (n-1) \right) \,.
\end{equation}
This relation comes from a full higher-order calculation, so higher-order 
expressions can also be used if desired to compute $n$ and $n_{{\rm T}}$ 
from a given inflation model, along the lines of Refs.~\cite{SL,CKLL2}.
Using this, $r$ is related to the spectral indices by
\begin{equation}
\label{nTr}
r = -6.2 n_{{\rm T}}\left( 1 - \frac{n_{{\rm T}}}{2} + (n-1) \right) \,.
\end{equation}

We specify the normalization of the density perturbations, following 
Ref.~\cite{BW}, at the present
Hubble scale $k = a_0 H_0$,\footnote{It is irrelevant that
this differs from the scale at which we evaluated the spectra to
do the normalization. The two scales are simply related,
$A_{{\rm S}}^2(a_0 H_0) = 7^{1-n} A_{{\rm S}}^2(7a_0 H_0)$.} and 
define
\begin{equation}
\delta_{\rm H} \equiv A_{{\rm S}} (a_0 H_0) \,.
\end{equation}
By focusing on the normalization at such a large scale, we obtain a
result which is independent, to excellent accuracy, of the choice of
cosmological parameters, such as the present Hubble constant and the
nature of the dark matter.\footnote{For the record, we take pure cold
dark matter with $h=0.75$ and $\Omega_{{\rm baryon}}h^2 = 0.0125$.} The
one exception is a nonzero cosmological constant, which does
affect the large-angle anisotropies; we generalize our results to
that case in the Appendix.

Our main result is a fitting function for the COBE normalization, which is 
accurately represented by
\begin{equation}
\label{answer}
\delta_{{\rm H}}(n,r) = 1.91 \times 10^{-5} \; 
	\frac{\exp \left[ 1.01 (1-n) \right]}{\sqrt{1+ 0.75 r}} \;. 
\end{equation}
The $1\sigma$ observational error is 7\%. Within the region $0.7 \leq
n \leq 1.3$ and $-0.3 \leq n_{{\rm T}} \leq 0$ (the latter corresponding to 
$0 \leq r \lesssim 2$), the fit is good to within 1.5\% everywhere.  The 
change from varying other cosmological parameters (except the density 
parameter) is within 4\% for reasonable variations.  In addition,
there is a systematic uncertainty of $\sim3\%$ associated with the
process by which the {\sl COBE} maps are made. Combining all of these 
uncertainties in quadrature, we believe that a realistic
estimate of the uncertainty in $\delta_{{\rm H}}$ is 9\% at $1\sigma$.

The terms in Eq.~(\ref{answer}) have a simple interpretation. The
numerical prefactor is the result for a scale-invariant density
perturbation spectrum. The $n$ term represents the pivot point of the COBE
data; it guarantees that COBE normalized spectra at fixed $r$ cross at 
$k_{{\rm pivot}} = e^{2.02} a_0 H_0$. This number actually corresponds to 
about the fifteenth multipole (and in fact a purely scalar fit even prefers 
the sixteenth), but the tensors give greater weight to the lower multipoles,
making $\ell=14$ a better overall choice for the pivot. The pivot point is 
at higher $\ell$ in the four-year data than in the two-year data, since the 
low multipoles were already cosmic-variance limited and the higher $\ell$ 
have improved signal-to-noise ratio. 

The final term in Eq.~(\ref{answer}) is the reduction in amplitude due to 
the tensors. It is interesting that the tensor term has a coefficient of 
only 0.75, since the definition of $r$ was intended to make that coefficient 
close to unity. However, the factor 12.4 in Eq.~(\ref{tensca}) was computed 
\cite{LL} using the fully matter-dominated Sachs-Wolfe approximation for 
both the tensor and scalar spectra. The dominant correction to this is the 
effect on the tensor spectrum from the universe not being perfectly matter 
dominated at last scattering \cite{TurWhi} (about twenty percent), with 
the start of the rise to the acoustic peak from the density perturbations 
contributing another five percent. These corrections have been noted in 
papers concerned with `cosmic confusion' \cite{confusion}. That the 
coefficient is only 0.75 means that papers using the original analytic 
argument have somewhat over-estimated the amount by which tensors reduce the 
power spectrum normalization.

\subsection{The inflationary energy scale}

For a given inflationary model, the COBE normalization fixes the
energy scale of inflation at that time, $V_*$. The fitting function
can therefore be inverted to supply the inflationary energy scale
\cite{engscal,TurWhi}. One would like to use Eq.~(\ref{dp}) in order to do 
this, but in fact this equation is not always very accurate at determining 
the amplitude, since the slow-roll approximation is not necessarily as 
accurate as the observations and fits we have discussed. We therefore use 
the next-order version of this equation, as derived by Stewart and Lyth
\cite{SL}, which is\footnote{To derive this from their results, some further 
relations are necessary which can be found in Ref.~\cite{LPB}.} 
\begin{equation}
A_{{\rm S}}^2(k) = [1 + 4.0 \epsilon - 2.1 \eta] \,
	\frac{512 \pi}{75} \, \left. 
	\frac{V^3}{m_{{\rm Pl}}^6 V'^2} \right|_{k = aH} \,.
\end{equation}
It is vital to note that we did not have to use this for the
normalization, because the amplitude correction in the prefactor is 
typically almost constant across the COBE scales (that is, $\epsilon$ and 
$\eta$ hardly vary). It therefore cancels out when one does the COBE 
normalization for large-scale structure.

We specify the energy at $k_*$, the place where the slow-roll
parameters were evaluated. Since $A^2(k_*) = 7^{n-1} \delta_{{\rm H}}^2$, 
substituting in the fitting function and carrying out a small parameter 
expansion gives
\begin{eqnarray}
\label{energy}
V_*^{1/4} &=& (5.4 \times 10^{-3} \, m_{\rm Pl}) \, \epsilon_*^{1/4} \, 
          \left( 1-3.2 \epsilon_* + 0.5 \eta_* \right)\,,
	\\ \nonumber
	&=& (6.6 \times 10^{16} \, {\rm GeV}) \, \epsilon_*^{1/4} \, 
          \left( 1-3.2 \epsilon_* + 0.5 \eta_* \right) \,.
\end{eqnarray}
Even if one does not have a specific inflation model in mind, it 
is possible to use this to obtain an upper bound on the energy density at 
the end of
inflation by imposing some assumptions about how inflation will end 
\cite{infeng}.

\subsection{On the validity of the power-law approximation}

It is possible to test the validity of approximating the spectra by
power-laws. We'll concentrate on the density perturbations. To a first
approximation, variation in the spectral indices is driven by the
difference between $n-1$ and $n_{{\rm T}}$, but one should consider
the full slow-roll formula for that variation at $k_*$ which is
\cite{KT}
\begin{equation}
\left. \frac{dn}{d \ln k} \right|_* = -24 \epsilon_*^2 + 16
	\epsilon_* \eta_* - \frac{m_{{\rm Pl}}^4}{32 \pi} 
	\frac{V'_* V'''_*}{V_*^2} \,.
\end{equation}
The COBE data extend only for slightly more than one log interval in either
direction about the central point, so unless this number is greater than, 
say, a few hundredths, the power-law approximation for COBE will be 
excellent. Kosowsky and Turner \cite{KT} evaluated this for a range of 
inflation models without finding a value anywhere near this large.

It is possible that the spectra may be well approximated by
power-laws on the COBE scales, but not across the much wider range
corresponding to future microwave anisotropy observations and to
large-scale structure \cite{HSSW,KT}. In that case, one simply uses our
results with the appropriate approximate power-law at the COBE scales.

\section{A worked example: the quadratic potential}

The simplest inflationary model is chaotic inflation \cite{chaotic} with a 
quadratic potential $V(\phi) = m^2 \phi^2/2$. For this potential, the 
slow-roll parameters are
\begin{equation}
\epsilon = \eta = \frac{1}{4\pi} \frac{m_{{\rm Pl}}^2}{\phi^2} \,,
\end{equation}
and hence $\phi_{{\rm end}} \simeq m_{{\rm Pl}}/\sqrt{4\pi}$. For 
definiteness, we take $N_* = 60$, and from Eq.~(\ref{efold}) we find $\phi_* 
= 3.10 \, m_{{\rm Pl}}$. So for this model one predicts
\begin{equation}
n = 0.967 \quad ; \quad n_{{\rm T}} = - 0.017 \quad ; \quad r = 0.10 \;.
\end{equation}

The normalization of the matter power spectrum, from Eq.~(\ref{answer}), is 
therefore $\delta_{{\rm H}} = 1.91 \times 10^{-5}$. Although this model is 
very close to scale-invariant limit, the difference is still non-negligible. 
For example, if one computes the variance at $8 h^{-1}$ Mpc, denoted 
$\sigma_8$, it is reduced, relative to scale-invariance with no 
gravitational waves, by 10\%, where $6\%$ is due to the tilt and $4\%$ due 
to the gravitational waves. Many other inflation models give much larger 
corrections than this.

The inflationary energy scale corresponding to this, from
Eq.~(\ref{energy}), is
\begin{equation}
V_*^{1/4} = 1.6 \times 10^{-3} \, m_{{\rm Pl}} = 2.0 \times 10^{16} \,
	{\rm GeV} \,,
\end{equation}
corresponding to $m = 1.2 \times 10^{-6} \, m_{{\rm Pl}}$. 

Finally, the scale-dependence of the spectral index is \mbox{$dn/d \ln k|_* 
= -5 \times 10^{-4}$}, which is completely negligible for COBE.

\section{Summary}

One of the most important uses of the COBE data is to normalize the matter 
power spectrum used in large-scale structure studies. We have shown that it 
is possible to condense this information into a single fitting function, 
covering not just the case of tilted perturbation spectra but also including 
the spectrum of gravitational waves that inflation predicts. A further 
extension to models with a cosmological constant is given in the 
Appendix. Within the slow-roll paradigm, the amount of gravitational waves, 
parameterized by $r$, is completely independent of the density perturbation 
slope $n$, though the scale-dependence of the gravitational wave spectrum is 
then predicted \cite{LL,LLrep}. The fitting functions can also be used for 
the case without gravitational waves, simply by setting $r$ equal to zero.

%%%%%%%%%%%%%%%%%%%%%%%%%%%%%%%%%%%%%%%%%%%%%%%%%%%%%%%%%%%%%%%%%%
\section*{Acknowledgments}

E.F.B. was supported by NASA, A.R.L. by the Royal Society and M.W. by the 
NSF and the DOE. We thank Kris G\'{o}rski for useful discussions, and Pedro 
Viana for spotting an error in the original version of Eq.~(\ref{superfit}). 
A.R.L. thanks Fermilab its hospitality during the initial stages of this 
work.

%%%%%%%%%%%%%%%%%%%%%%%%%%%%%%%%%%%%%%%%%%%%%%%%%%%%%%%%%%%%%%%%%%
\appendix

\section{Generalization to models with a cosmological constant}

The dynamics of inflation are insensitive to whether or not there is a
present cosmological constant $\Lambda$, provided the spatial geometry
is kept flat.  We now provide the generalization of our expressions to
the case where $\Lambda\ne 0$.

We need a generalization of Eq.~(\ref{powerspec}) to account for the 
change in the growth of perturbations in a low-density universe, and the 
relation between the curvature perturbation and the matter power spectrum.
If $\delta_{{\rm H}}$ continues to indicate the present power spectrum,
and $A_{{\rm S}}$ the initial perturbation spectrum as before, then
\begin{equation}
\delta_{{\rm H}} \equiv \frac{g(\Omega_0)}{\Omega_0} 
	A_{{\rm S}} (a_0 H_0) \,,
\end{equation}
and the right hand side of Eq.~(\ref{powerspec}) is multiplied by 
$g^2(\Omega_0)/\Omega_0^2$.  Here $g(\Omega)$ is the growth suppression 
factor, which is accurately fit by \cite{CPT}
\begin{equation}
g(\Omega) = \frac{5}{2} \Omega \left[ \frac{1}{70} + 
	\frac{209 \Omega}{140} - \frac{\Omega^2}{140} + 
	\Omega^{4/7} \right]^{-1} \,.
\end{equation}

A generalization of the factor 0.75 multiplying the gravitational wave term 
in the fitting function for $\delta_{{\rm H}}$ is also needed. By directly 
evaluating the radiation power spectra for scale-invariant initial spectra 
with $A_{{\rm S}} = A_{{\rm T}} = 1$, we find the ratio of contributions to 
$C_{14}$ is fit to within a percent by
\begin{equation}
\label{fom}
\frac{C_{14}^{{\rm T}}}{C_{14}^{{\rm S}}} \equiv f(\Omega_0) \simeq 
	0.75 - 0.13 \, \Omega_\Lambda^2 \,,
\end{equation}
with $\Omega_\Lambda=1-\Omega_0$.

First considering the case with no tensors, a good fit is obtained by using 
an $\Omega_0$ dependence plus a cross-term between $\widetilde{n} = n-1$ and 
$\Omega_\Lambda$.
The formula
\begin{eqnarray}
\label{lamanswer}
\delta_{{\rm H}}(n,\Omega_0) & = & 1.91 \times 10^{-5} \, 
	\exp \left[-1.01 \, \widetilde{n} \right] \, 
	\Omega_0^{-0.80 - 0.05 \ln \Omega_0} \nonumber \\
	& & \hspace*{24pt} \times \left[ 1 + 
	0.18 \, \widetilde{n} \, \Omega_\Lambda \right] \,, 
\end{eqnarray}
holds within 2.5\% for $n$ as before and $0.2 \leq \Omega_0 \leq 1$.

We can extend this to include tensors using Eq.~(\ref{fom}) and the addition 
of an extra cross-term which vanishs in the critical-density case and the 
tensorless case. This gives
\begin{eqnarray}
\label{superfit}
\delta_{{\rm H}} & = & 1.91 \times 10^{-5} \,
	\frac{\exp \left[ -1.01 \, \widetilde{n} \right]}{\sqrt{1+ 
	f(\Omega_0) \, r}}  \; \Omega_0^{-0.80 - 0.05 \ln \Omega_0} 
\nonumber \\
 & & \hspace*{24pt} \times \left[ 1 + 0.18 \, \widetilde{n} \,
 	\Omega_\Lambda - 0.03 \, r \, \Omega_\Lambda \right]  \,. 
\end{eqnarray}
For the parameter ranges $0.7\leq n\leq 1.3$, $-0.3\leq n_{{\rm T}}\leq 0$
and $0.2 \leq \Omega_0 \leq 1$, this fit is within 1\% almost everywhere, 
and always within 2.5\%. The relation between $n_{{\rm T}}$ 
and $r$ is still given by Eq.~(\ref{nTr}).

%%%%%%%%%%%%%%%%%%%%%%%%%%%%%%%%%%%%%%%%%%%%%%%%%%%%%%%%%%%%%%%%%%

\end{document}